\documentclass[prd,twocolumn,showpacs,preprintnumbers,superscriptaddress]{revtex4}
\usepackage{hyperref,amssymb,amsmath,mathrsfs,bm,graphicx}
\usepackage{color}
\begin{document}
\title{Nonadiabatic charged spherical evolution in the 
postquasistatic approximation}
\author{L. Rosales}
\affiliation{Laboratorio de F\'\i sica Computacional,
Universidad Experimental Polit\'ecnica ``Antonio Jos\'e de
Sucre'', Puerto Ordaz, Venezuela}
\author{ W. Barreto}
\affiliation{Centro de F\'\i sica Fundamental, Facultad de Ciencias, Universidad de Los Andes, M\'erida, Venezuela\footnote{On sabbatical leave at Universitat de les Illes Balears, Palma de Mallorca, Spain.}}
\author{C. Peralta}
\affiliation{Deutscher Wetterdienst, Frankfurter Str. 135, 63067 Offenbach, Germany}
\affiliation{School of Physics, University of Melbourne, Parkville,
VIC 3010, Australia}
\author{B. Rodr\'\i guez--Mueller}
\affiliation{Computational Science Research Center, College of Sciences,
San Diego State University, San Diego, California, USA}
\begin{abstract}
We apply the postquasistatic approximation, an iterative
method for the evolution of self--gravitating spheres of matter,
to study the evolution of dissipative and electrically charged distributions in General Relativity. {The numerical implementation of our approach leads to a solver which is globally second--order convergent}.
We evolve nonadiabatic distributions assuming an equation
of state that accounts for the anisotropy induced by the
electric charge.
Dissipation is described by streaming out or diffusion approximations. We match the interior solution, in noncomoving coordinates, with the Vaidya--Reissner--Nordstr\"om exterior solution. 
Two models are considered: i) a Schwarzschild--like shell in the diffusion limit; ii) a Schwarzschild--like interior in the free streaming limit. These toy models tell us something about the nature
of the dissipative and electrically charged collapse. Diffusion stabilizes the gravitational collapse producing a spherical shell whose contraction is halted in a short characteristic hydrodynamic time. 
The streaming out radiation provides a more efficient mechanism for emission of energy,
redistributing the electric charge on the whole sphere, while the distribution collapses indefinitely with a longer hydrodynamic time scale.
\end{abstract}
\maketitle
\section{Introduction}
{Despite their apparent simplicity, 1+1 models of the fluid dynamics of compact objects in numerical relativity can include realistic transport mechanisms and equations of state.
{Renewed interest on electric charge in stars has driven the numerical
integration of the Einstein--Maxwell (EM) system. Current integrators
of the EM system are in comoving coordinates \cite{ms64}, and seem to be
limited to one dimensional numerical solvers \cite{g05}, \cite{gl07} of the May and
White family \cite{mw66}, \cite{mw67}.}
Because of the obvious interest in three--dimensional situations, it is desirable to
use noncomoving coordinates. Numerical simulations
to explore the relevance of electric charge in the process
of dissipative and anisotropic (viscous) gravitational collapse
are desirable as well.}

{The numerical solution of Einstein equations  in 3+1 dimensions is an essential tool
for the investigation of strong field scenarios of astrophysical interest (see \cite{lehner} and references therein). Numerical relativity has led to the discovery of critical phenomena in gravitational collapse \cite{choptuik}, allowed the study of binary black holes and neutron stars \cite{pretorius}--\cite{bgr08} and the development of relativistic hydrodynamics solvers \cite{font}, among other major achievements.
The main limitation currently faced by realistic models in numerical
relativity is the computational demand in three--dimensional evolution. 
Computationally less intensive 1D models still remain an interesting
alternative and help narrow the search in the parameter space for
general solvers.
These simplified systems provide a necessary test bed to study the phenomena expected in fully realistic three--dimensional configurations.}

In this paper, we study a self--gravitating spherical distribution of charged matter containing a dissipative fluid. We use noncomoving coordinates and follow the method reported in \cite{brm02}. Herrera et al. realized that this method was equivalent to going one step further from the quasistatic regime, and consequently has been named the postquasistatic approximation (PQSA) after \cite{hbds02}. 

The essence of the PQSA was first proposed in \cite{hjr80}
using radiative Bondi coordinates and it has been extensively
used by Herrera and collaborators \cite{almostall}--\cite{hmnp}. In the
context of charged distributions of matter the original method was used 
as well \cite{mnrp88}, \cite{bd96}, \cite{pr96}.
The approach is based on the introduction of a set of
conveniently defined ``effective" variables, which are the effective
pressure and energy density, and a heuristic ansatz on
the latter \cite{brm02}.
By QSA we mean that the effective
variables coincide with the corresponding physical variables
(pressure and energy density). In Bondi
coordinates the notion of QSA is
not evident: the system goes directly from static to
postquasistatic evolution. In an adiabatic and slow
evolution we can catch--up that phase, clearly seen in
noncomoving coordinates; this can be achieved using
Schwarzschild coordinates \cite{hbds02}. 
If the configuration is leaving equilibrium, the PQSA
description seems to be
enough and can be used as a test bed in numerical relativity  \cite{b09}.
Its systematic use of local Minkowskian and comoving observers, named Bondians, was used to reveal a central
equation of state in adiabatic scenarios \cite{bcb09}, and to couple matter with radiation \cite{bcb10}.

In self--gravitating systems the electric charge is believed
to be constrained by the fact that the resulting electric
field should not exceed the critical field for pair creation,
$10^{16}$ V cm$^{-1}$ \cite{b71}. This restriction in the critical  field
has been questioned \cite{rebels1}--\cite{rebels4} and does not apply to phases of
intense dynamical activity with time scales of the order
of (or even smaller than) the hydrostatic time scale, and
for which the QSA \cite{hd97}
is clearly not reliable as in the collapse of very massive stars or the quick
collapse phase preceding neutron star formation (see \cite{dhlms07} and
references therein). {Electric charge has been also studied mostly under static conditions \cite{dhlms07}, \cite{i02}, \cite{varela}, \cite{remlz03}.
It is of recent interest the charged quasiblack holes \cite{lemos} and its extension
to quasispherical realization \cite{bonnor}.}
{Distributions electrically charged can be considered in practice as anisotropic \cite{brrs07}, \cite{ivanov}.
Authors combine anisotropy and electric charge \cite{dhlms07}, \cite{maharaj}, \cite{ma10}
but not as a single entity by means of an equation of state.}

The electric field has been postulated to be very high in strange
stars with quark matter \cite{u04}, \cite{mh04}, although other authors suggest
that strange stars wouldn't need a large electrical field \cite{jrs06}.
The effects of dissipation, in both limiting cases
of radiative transport, within the context of the QSA,
have been studied in \cite{hs03}. Using this
approximation is very sensible because the hydrostatic
time scale is very small, compared with stellar lifetimes,
for many phases of the life of a star. It is of the order
of 27 minutes for the sun, 4.5 seconds for a white dwarf
and 104 seconds for a neutron star of one solar mass
and 10 Km radius \cite{books1}--\cite{books2}. However, such an approximation
does not apply to the very dynamic phases mentioned
before. In those cases it is mandatory to take into account
terms which describe departure from equilibrium,
i.e. a full dynamic description has to be used \cite{hs04}.

In this paper
we consider that the electric charge can be seen as anisotropy \cite{brrs07}, but not any anisotropy as we shall see.
For certain density ranges, locally anisotropic pressure can be physically justified
in self--gravitating systems, since different kinds of physical phenomena
may take place, giving rise to local anisotropy and in turn relaxing the upper
limits imposed on the maximum value of the surface gravitational potential 
\cite{l33}.
The influence of local anisotropy in general relativity has been studied mostly
under static conditions (see \cite{hs97} and references therein; \cite{ema07}). Herrera et al. \cite{hdmost04} have reported a general study for spherically symmetric dissipative anisotropic fluids
with emphasis on the relationship among the Weyl tensor, the shear tensor,
the anisotropy of the pressure and the density inhomogeneity.

On the other hand, it is well known that different energy--momentum tensors can lead to the same spacetime \cite{equivalents1}--\cite{equivalents5}.  For instance,
viscosity can be considered as a special case of anisotropy \cite{visco}. Here we illustrate this idea for the Einstein--Maxwell system under spherical symmetry. To accomplish that program we use the total energy characterization as in \cite{b71} and \cite{mp79}. The electric energy (or pressure) contributes to the fluid in a such way that the electrically charged perfect fluid is equivalent to an anisotropic fluid under certain conditions.

{Massive stars evolve emitting massless particles (photons and/or neutrinos).
Neutrino emission seems to be the only plausible mechanism to carry away
the bulk of binding energy of a collapsing star, leading to a black hole
or neutron star \cite{ks79}. It seems clear that the free--streaming process 
is associated with the initial stages of the collapse, 
while the diffusion approximation becomes valid toward
the final stages.
Observation from supernova 1987A indicates that the regime of radiation
transport prevailing during the emission process is closer to the
diffusion approximation than to the free streaming \cite{l88}. 
Heat flow is usually considered as proportional to the gradient of temperature. This
assumption is very sensible because the mean free path of particles responsible for the propagation of energy in stellar interiors is very small as compared with the typical length of the object \cite{hd97}.}

{Although some transport equations in the relaxation time approximation
have been proposed (see for instance \cite{m96} and
references therein), the evolution of temperature profiles in the context
of general relativity remains an unsolved problem. Therefore,
we avoid stating any explicit evolution equation for
heat flow in this investigation. Recently, some progress
have been  achieved by Herrera and collaborators on the
study of dissipation via an appropriate causal procedure 
(see for example  \cite{hdhmm97,hm98,hdmost04,hdb06}). In the present investigation
 we obtain the zeroth order level of approximation for heat
flow profiles, which will serve as the basis of a future investigation
which includes dissipation in a realistic way using the
M\"uller--Israel--Stewart theory \cite{m67,i76,is76,is79}.
We already see an important advantage using the PQSA studying
configurations with anisotropy (induced by shear viscosity), streaming
out and heat flow processes in spherical collapse: an observer using
radiating coordinates to study heat flow misses some important
details.

The paper is organized as follows. In Sec. II we write the field equations for Bondian observers to show how the electric charge induces anisotropy, matching with the exterior Reissner--Nordstr\"om--Vaidya solution, and write the surface equation, following the PQSA protocol \cite{hbds02}, \cite{prrb10}. 
In Sec. III we present a summary of the numerical methods employed.
In Sec. IV we {show local and global tests of numerical convergence and} illustrate the PQSA integration procedure with two
nonadiabatic charged models. In Sec. V we conclude with some remarks.

\section{The Einstein--Maxwell system}
\subsection{Field equations for Bondian observers}
~~~~To write the Einstein field equations we use the line element in Schwarzschild--like coordinates
\begin{equation}
ds^2=e^\nu dt^2-e^\lambda dr^2-r^2\left( d\theta ^2+\sin^2\theta d\phi ^2\right), \label{eq:metric} 
\end{equation}
where $\nu = \nu(t,r)$ and $\lambda = \lambda(t,r)$, with
 $(t,r,\theta,\phi)\equiv(0,1,2,3)$.
 In order to get physical input we introduce the
 Minkowski coordinates $(\tau,x,y,z)$ by \cite{b64}
\begin{equation}
d\tau=e^{\nu /2}dt,\,  
dx=e^{\lambda /2}dr,\,  
dy=rd\theta,\, 
dz=r \sin \theta d\phi.\label{eq:local}
\end{equation}
In these expressions $\nu$ and $\lambda$ are constants, because they
 have only local values. 
Next we assume that, for an observer moving relative to these coordinates
 with velocity $\omega$ in the radial direction, the space contains
 a nonstatic distribution of matter which is spherically
symmetric and consists of charged fluid of energy density $\rho$, pressure $p$,
electric energy density $\rho_e$, radiation energy flux $q$ diffusing in
the radial direction, and unpolarized radiation of energy density $\epsilon$.
Thus, the  energy--momentum tensor is 
\begin{equation}
T_{\mu\nu}=(\rho+p)u_{\mu}u_{\nu}-pg_{\mu\nu}+\epsilon l_\mu l_\nu+q_{\mu}u_{\nu}+ q_{\nu}u_{\mu}+E_{\mu\nu},
\end{equation}
where $u_{\alpha}$,  $l_\alpha$, $q_{\alpha}$ are the 4--velocity, the 4--null vector
and the heat flux 4--vector respectively, which  satisfy
$\,\,\,\,u^\alpha u_\alpha=1$, $q_{\alpha}u^{\alpha}=0$, $l^\alpha l_\alpha=0$, and 
$E_{\mu\nu}$ is the electromagnetic energy--momentum tensor 
\begin{equation}
E_{\mu\nu}=\frac{\pi}{4}\left[F_\mu^{\,\,\kappa}F_{\nu\kappa}+\frac{1}{4}g_{\mu\nu}F_{\sigma\kappa}F^{\sigma\kappa}\right],
\end{equation}
where $F_{\mu\kappa}$ is the Maxwell field tensor,  which satisfies the Maxwell equations:

\begin{equation}
F_{[\mu\nu;\sigma]}=0 
\end{equation}
and
\begin{equation}
(\sqrt{-g}F^{\mu\nu})_{,\nu}=4\pi\sqrt{-g}J^{\mu},
\end{equation}
where the semicolon (;) and the comma (,) represent covariant derivative and partial differentiation with 
 respect to the indicated coordinate, respectively; $J^{\mu}=\sigma u^{\mu}$ is
electric  current 4-vector and $\sigma$ the electric conductivity. Because of the spherical symmetry, only the radial electric field
$F^{tr}=-F^{rt}$ is nonzero. On the other hand, the inhomogeneous Maxwell equations become
\begin{equation}
s_{,r}=4\pi r^{2}J^{t}e^{\frac{1}{2}(\nu+\lambda)}
\end{equation}
and
\begin{equation}
s_{,t}=-4\pi r^{2}J^{r}e^{\frac{1}{2}(\nu+\lambda)},
\end{equation}
where $J^{t}$  and $ J^{r}$  are the temporal and radial components of the current 4-vector, respectively. The function $s(t,r)$ is naturally interpreted
as the charge within the radius $r$ at the time $t$. We define
the function $s(t,r)$ by $F^{tr}=se^{-(\lambda+\nu)/2}/r^{2}$,
with 
\begin{equation}
s(t,r)=\int4\pi r^{2}J^{t}e^{-(\lambda+\nu)/2}dr.
\end{equation} 

The conservation of charge inside a sphere comoving with the fluid is expressed as
\begin{equation}
u^{\alpha}s_{,\alpha} =0.      
\end{equation}
We can write the conservation equation in a more suitable form for numerical
purposes
\begin{equation}
s_{,t}+\frac{dr}{dt}s_{,r}=0, \label{eq:ce} 
\end{equation}
where the velocity of matter in the Schwarzschild coordinates is
\begin{equation}
\frac{dr}{dt} = \omega e^{(\nu-\lambda)/2}. \label{eq:velocity}
\end{equation}

The contravariant components of the 4--velocity, the heat flux 4--vector and the null outgoing vector are
\begin{equation}
u^{\mu}=\frac{e^{-\nu/2}}{(1-\omega^{2})^{1/2}}\delta_{t}^{\mu}+\frac{\omega e^{-\lambda/2}}{(1-\omega^{2})^{1/2}}\delta_{r}^{\mu},
\end{equation}
\begin{equation}
q^{\mu}=\frac{\omega e^{-\nu/2}q}{(1-\omega^{2})^{1/2}}\delta_{t}^{\mu}+\frac{e^{-\lambda/2}q}{(1-\omega^{2})^{1/2}}\delta_{r}^{\mu}
\end{equation}
and
\begin{equation}
l^\mu=e^{-\nu/2}\delta_{t}^{\mu}+e^{-\lambda/2}\delta_{r}^{\mu}.
\end{equation}
We write the field equations for the Einstein--Maxwell system in relativistic units ($G=c=1$) as follows: 
\begin{eqnarray}
&&\frac{\rho+p\omega^{2}}{1-\omega^{2}}+\frac{2\omega q}{1-\omega^{2}}+
 \frac{1+\omega}{1-\omega}\epsilon+\frac{s^{2}}{8\pi r^{4}} =\nonumber \\
&&\frac{1}{8\pi r}\left[\frac{1}{r} - 
e^{-\lambda}\left(\frac 1{r}-\lambda_{,r}\right)\right], \label{eq:ee1}
\end{eqnarray}

\begin{eqnarray}
&&\frac{p+\rho\omega^{2}}{1-\omega^{2}}+\frac{2\omega q}{1-\omega^{2}}+ \frac{1+\omega}{1-\omega}\epsilon-\frac{s^{2}}{8\pi r^{4}} =
\nonumber\\
&&\frac{1}{8\pi r}\left[
e^{-\lambda}\left(\frac 1{r}+\nu_{,r}\right) - \frac{1}{r}\right], \label{eq:ee2}
\end{eqnarray}

\begin{eqnarray}
&&p + \frac{s^{2}}{8\pi r^{4}}=\nonumber\\
&& \frac{1}{32\pi} \{ e^{-\lambda}[ 2\nu_{,rr}+\nu_{,r}^2
-\lambda_{,r}\nu_{,r} + \frac{2}{r}
(\nu_{,r}-\lambda_{,r}) ]\nonumber \\ \nonumber \\
&&-e^{-\nu}[ 2\lambda _{,tt}+\lambda_{,t}(\lambda_{,t}-\nu_{,t}) ] \} \label{eq:ee3}
\end{eqnarray}
and
\begin{eqnarray}
&&\frac{\omega}{1-\omega^{2}}(p+\rho)+\frac{(1+\omega^{2})}{1-\omega^{2}}q +\frac{1+\omega}{1-\omega}\epsilon  = \nonumber \\
&&-\frac{\lambda_{,t}}{8\pi r}e^{-\frac 12(\nu+\lambda)}. \label{eq:ee4}
\end{eqnarray}

\subsection{Anisotropy induced by electric charge}

To write the field equations in a form equivalent to an anisotropic fluid,
we introduce 
\begin{equation}
e^{-\lambda}=1-2\mu/r,
\end{equation}
where
\begin{equation}
\mu(t,r)=m(t,r)-\frac{s^2}{2r},
\end{equation}
\begin{widetext}
$m$ being the total mass \cite{b71}, \cite{mp79}. Thus the field equations (\ref{eq:ee1})--(\ref{eq:ee4}) read
\begin{equation}
\tilde{\rho}=\frac{\mu_{,r}}{4\pi r^{2}},\label{eq:e1}
\end{equation}
\begin{equation}
\tilde{p}=\frac{1}{8\pi r}\left[\nu_{,r}(1-\frac{2\mu}{r})-\frac{2\mu}{r^2}\right],\label{eq:e2}
\end{equation}
\begin{eqnarray}
 p_{t}=\frac{(r-2\mu)}{16\pi r}\left\{\nu_{,rr}+\frac{\nu_{,r}^{2}}{2}+\frac{\nu_{,r}}{r}-\left(\nu_{,r}+\frac{2}{r}\right)\frac{(\mu_{,r}-\mu/r)}{(r-2\mu)}\right\} 
-\frac{e^{-\nu}}{8\pi (r-2\mu)}\left\{\mu_{,tt}+\frac{3\mu_{,t}^{2}}{(r-2\mu)}-\frac{\mu_{,t}\nu_{,t}}{2}\right\} \label{eq:e3}
\end{eqnarray}
\end{widetext}
and
\begin{equation}
S=-\frac{\mu_{,t}}{4\pi r}(1-2\mu/r)^{\frac{1}{2}}e^{-\frac{1}{2}\nu},\label{eq:e4}
\end{equation}
where the conservative variables are
\begin{equation}
\tilde{\rho}=\frac{\hat{\rho}+p_{r}\omega^{2}}{1-\omega^{2}}+
\frac{2\omega q}{1-\omega^{2}}
+\epsilon\frac{1+\omega}{1-\omega},
\end{equation}
\begin{equation}
S=\frac{\omega}{1-\omega^{2}}(p_{r}+\hat{\rho})+\frac{1+\omega^2}{1-\omega^2}q
+\epsilon\frac{1+\omega}{1-\omega},
\end{equation}
and the flux variable
\begin{equation}
\tilde{p}=\frac{p_{r}-\hat{\rho}\omega^{2}}{1-\omega^{2}}+
\frac{2\omega q}{1-\omega^{2}}
+
\epsilon\frac{1+\omega}{1-\omega},
\end{equation}
as in the standard ADM 3+1 formulation. Within the PQSA $\tilde\rho$ and $\tilde p$ are
referred as effective density and effective pressure, respectively.

Equations (\ref{eq:e1})--(\ref{eq:e4}) are formally the same as for
 an anisotropic fluid, with $\hat \rho=\rho + \rho_e$, $p_r=p-\rho_e$, $p_t=p+\rho_e$
 and the electric energy density $\rho_e=E^2/8\pi$, where $E=s/r^2$ is the local electric field intensity. If we define the degree of local anisotropy induced by charge as $\Delta=p_t-p_r=2\rho_e$, the electric charge determines such a degree at any point.

From (\ref{eq:e1}) and (\ref{eq:e4}) we easily obtain
\begin{equation}
\frac{d\mu}{dt}=-4\pi r^2\left\{\frac{dr}{dt}p_r
+\left[q +\epsilon(1+\omega)\right](1-2\mu/r)^{1/2}e^{\nu/2} \right\}. \label{eq:energy}
\end{equation}
This equation, known as the momentum constraint in the ADM 3+1 formulation,
expresses the power across any moving spherical shell.

It can be shown that
\begin{eqnarray}
&&\tilde p_{,r} + \frac{(\tilde\rho + \tilde p)(4\pi r^3\tilde p + \mu)}{r(r-2\mu)}
 + \frac{2}{r}(\tilde p - p_t)\nonumber \\
&=&\frac{e^{-\nu}}{4\pi r(r-2\mu)}\left( \mu_{,tt} +\frac{3\mu_{,t}^2}{r-2\mu}-
\frac{\mu_{,t}\nu_{,t}}{2}\right) .\label{eq:TOV}
\end{eqnarray}
This equation is the same as for an anisotropic fluid \cite{chew82} and is a  generalization of the hydrostatic support equation, that is, the Tolman--Oppenheimer--Volkoff (TOV) equation. Equation (\ref{eq:TOV}) is equivalent to the equation
of motion for the fluid in conservative form in the standard ADM 3+1 formulation \cite{b09}. 
Equation (\ref{eq:TOV}) leads to the third equation at the surface (see next section); up to this point
is completely general within spherical symmetry.

To close this sub--section we have to mention that we assume the following equation of state  (EoS) \cite{b93} for nonadiabatic modeling and only as initial--boundary datum:

\begin{equation}
p_t - p_r = \frac{C (\tilde{p} + \tilde{\rho})(4 \pi r^{3} \tilde{p} +
\mu)}{(r-2\mu)},
\label{eq:EDE}
\end{equation}
where $C$ is a constant.

\subsection{Junction conditions}
~~~~We describe the exterior spacetime by the Reissner--Nordstr\"om--Vaidya
metric
\begin{eqnarray}
ds_{+}^2&=&\left( 1-\frac{2\mathcal{M}(u)}r+\frac{{\ell}^2}{r^2}\right)
du^2+2du\,dr\nonumber\\
&-&r^2\left( d\theta ^2+\sin {}^2\theta \, d\phi ^2\right),
\end{eqnarray}
where $\mathcal{M}(u)$ is the total mass and $\ell$ the total charge, and $u$ is the retarded time. The exterior and interior solutions are separated by the surface $r=a(t)$.
 In order to match both regions on this surface we use the Darmois junction
 conditions. Demanding the continuity of the first fundamental form, we obtain
 \begin{equation}
e^{-\lambda_a}=1-\frac{2\mathcal{M}}{a}+\frac{\ell^2}{a^2},
\end{equation}
\begin{equation}
s_a=\ell,
\end{equation}
\begin{equation}
m_a(u)=\mathcal{M}
\end{equation}
and
\begin{equation}
\nu_a = -\lambda_a.\label{eq:ffb}
\end{equation}
 The subscript $a$ indicates that the quantity is evaluated 
 at the surface. In this work, we will use the continuity of the independent components of the energy--momentum flow instead of the second fundamental form, which have been shown to be equivalent \cite{hd97} and it is simpler to apply to the present case. 
 This last condition
 guarantees the absence of singular behaviors on the
 surface. It is easy to check that
\begin{equation}
p_{a}=q_{a} \label{eq:boundary},
\end{equation}
which expresses the continuity of the radial pressure across the boundary of the
distribution $r=a(t)$.
\subsection{Surface equations}
Following the protocol sketched in \cite{hbds02} we write the
surface equations evaluating (\ref{eq:velocity}), (\ref{eq:energy}) and (\ref{eq:TOV})
at the surface of the distribution. The first and second surface equations read
\begin{equation}
\frac{da}{dt}=\omega_a\left(1-\frac{2\mu_a}{a}\right),
\end{equation}
\begin{equation}
\frac{d\mu_a}{dt}=-L+\frac{\ell^2}{2a^2}\frac{da}{dt},
\end{equation}
with
\begin{equation}
L\equiv [Q +E(1+\omega_a)]\left(1-\frac{2\mu_a}{a}\right),
\end{equation}
where $E=4\pi a^2 \epsilon_a$ and $Q=4\pi a^2 q_a$.

We need a third surface equation to specify
 the dynamics completely for any set of initial conditions and a given
 luminosity profile $L(t)$. For this purpose we can use the field
 Eq. (\ref{eq:ee3}) or the conservation Eq. (\ref{eq:TOV}) written in terms of the effective variables, which is clearly model dependent.
\begin{figure}[htbp!]
\begin{center}
\scalebox{0.4}{\includegraphics[angle=0]{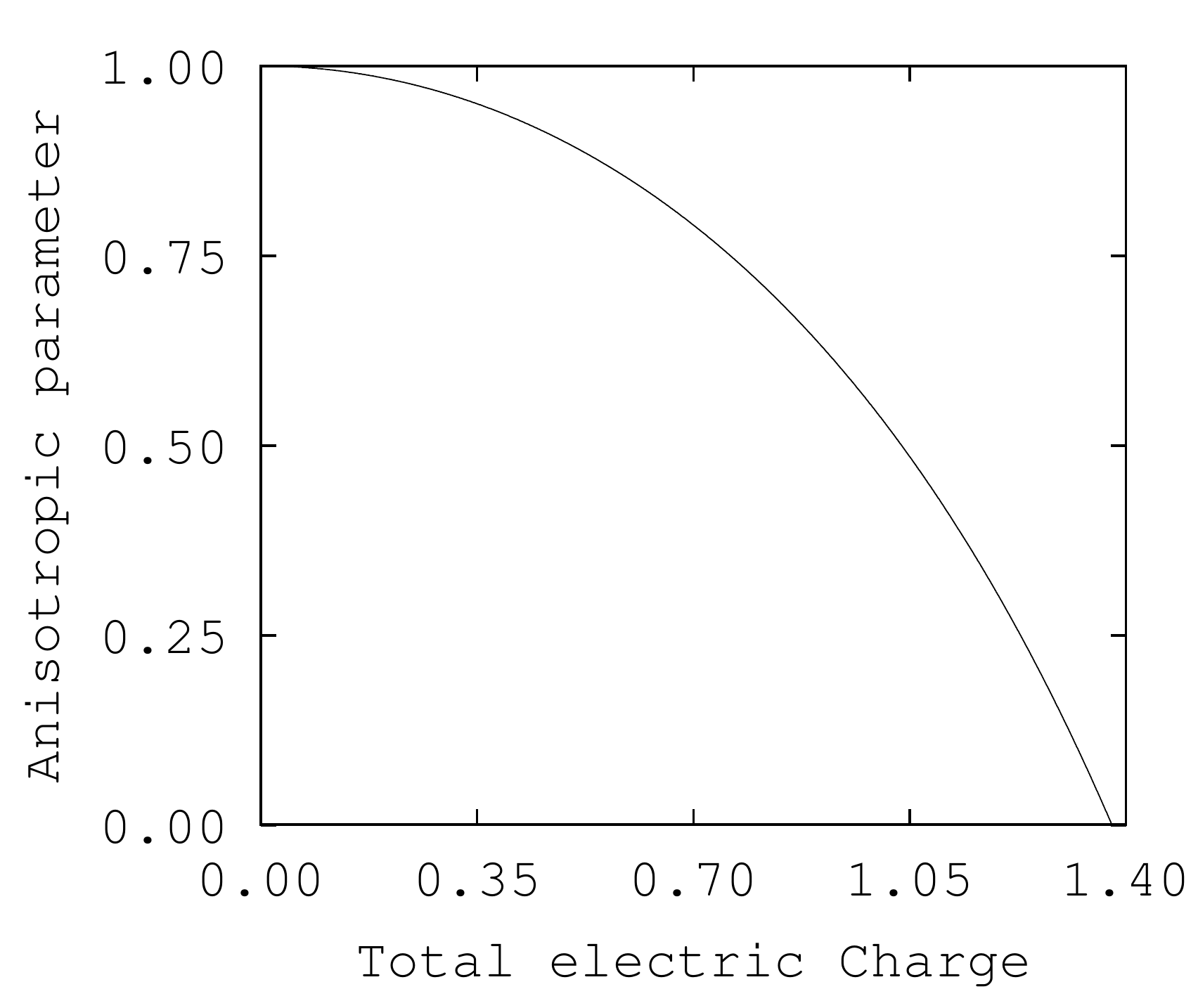}}
\caption{Anisotropic parameter $h$ as a function of the total electric charge $\ell$, calculated using (\ref{eq:EDE}) evaluated at the surface for the Schwarzschild--like model. The initial conditions are $a(0) = 5.0$, $m_a(0)=1.0$, $\omega_a(0)=-10^{-3}$.}
\end{center}
\label{fig:figure1}
\end{figure}
\begin{figure}[htbp!]
\begin{center}
\scalebox{0.4}{\includegraphics[angle=0]{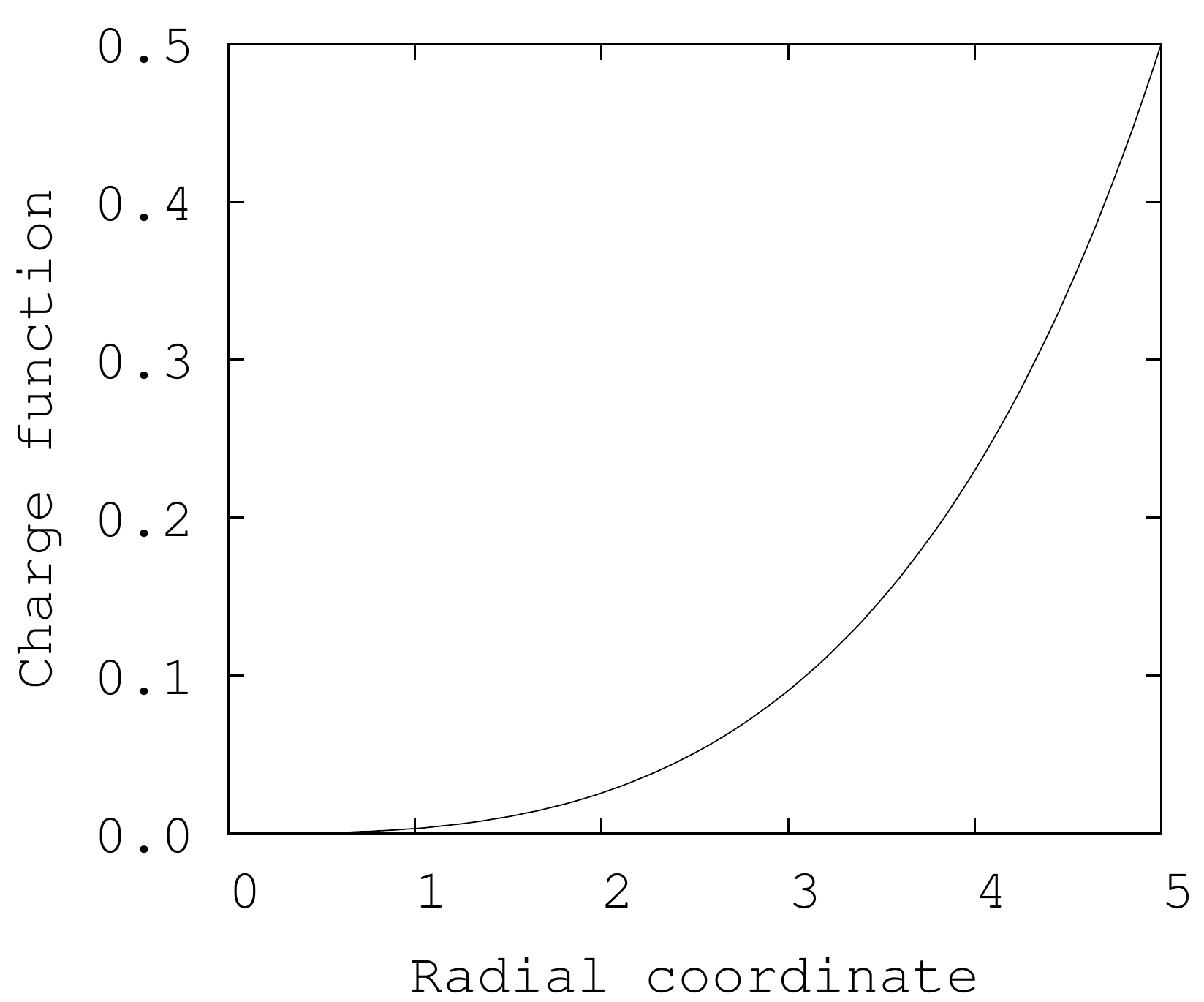}}
\caption{Initial profile of the charge function $s$, calculated using (\ref{eq:EDE}) for the Schwarzschild--like model. The initial conditions are $a(0) = 5.0$, $m_a(0)=1.0$, $\omega_a(0)=-10^{-3}$, with $\ell=0.5$, which corresponds to $h=0.8966$}
\end{center}
\label{fig:figure2}
\end{figure}

\begin{figure}[htbp!]
\begin{center}
\scalebox{0.4}{\includegraphics[angle=0]{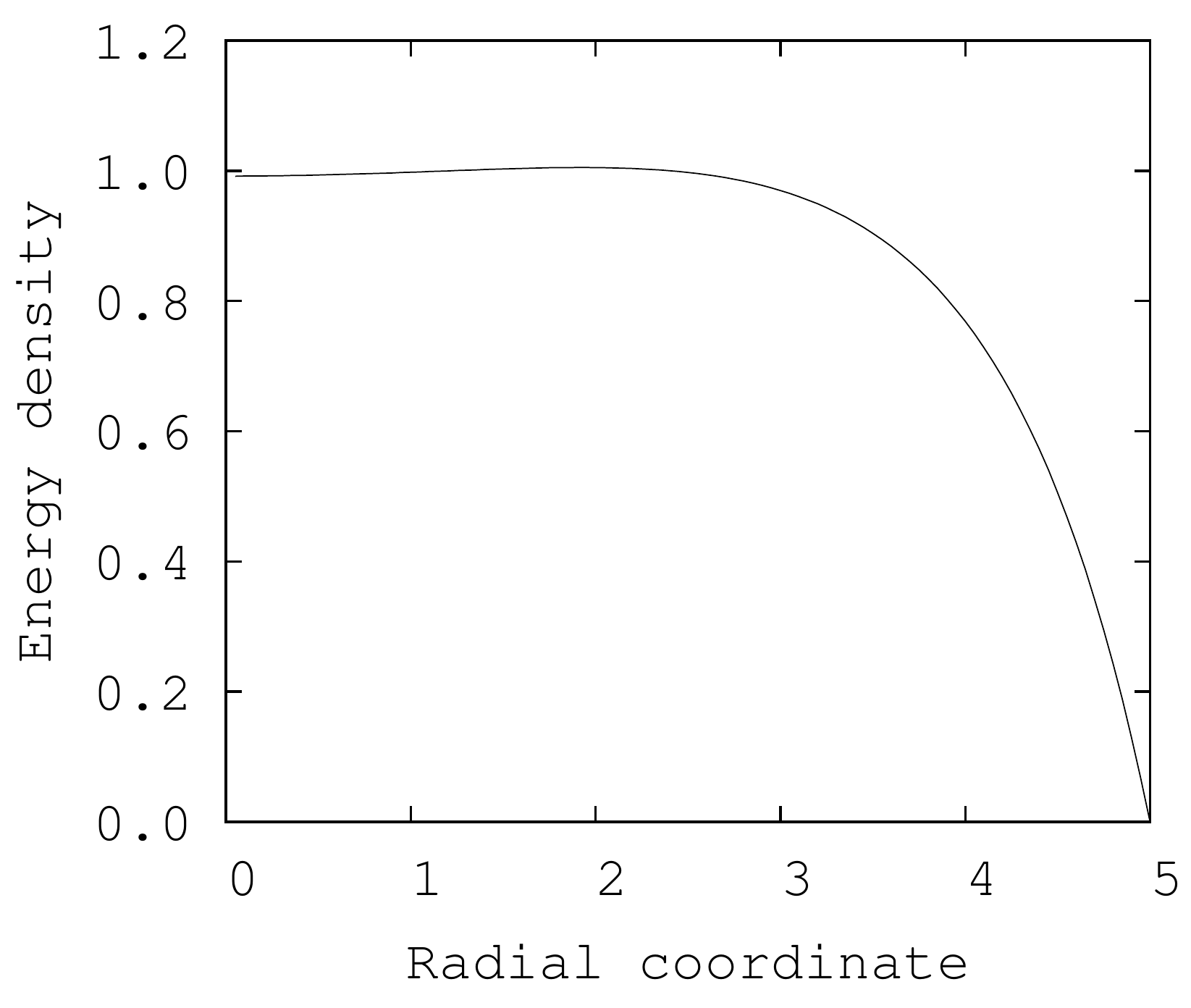}}
\caption{Initial profile of the energy density $\rho$ (a variation relative to the outermost value, multiplied by $10^6$) for the Schwarzschild--like model I. 
The initial conditions are $a(0) = 5.0$, $m_a(0)=1.0$, $\omega_a(0)=-10^{-3}$,
with $\ell=0.5$, which corresponds to $h=0.8966$.}
\end{center}
\label{fig:figure3}
\end{figure}

\begin{figure}[htbp!]
\begin{center}
\scalebox{0.4}{\includegraphics[angle=0]{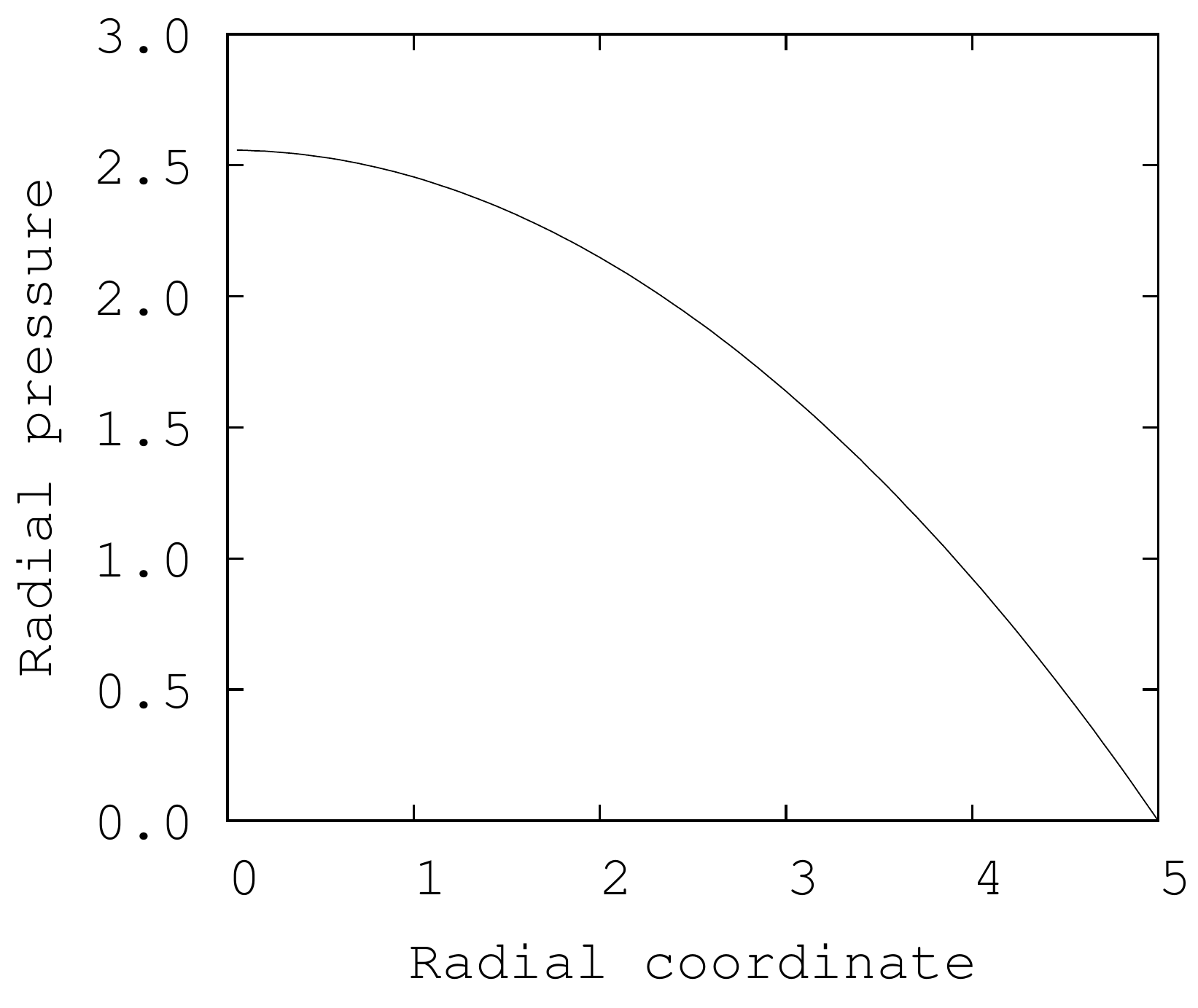}}
\caption{Initial profile of the radial pressure $p$ (multiplied by $10^4$) for the Schwarzschild--like model I. 
The initial conditions are $a(0) = 5.0$, $m_a(0)=1.0$, $\omega_a(0)=-10^{-3}$,
with $\ell=0.5$, which corresponds to $h=0.8966$.}
\end{center}
\label{fig:figure4}
\end{figure}

\begin{figure}[htbp!]
\begin{center}
\scalebox{0.4}{\includegraphics[angle=0]{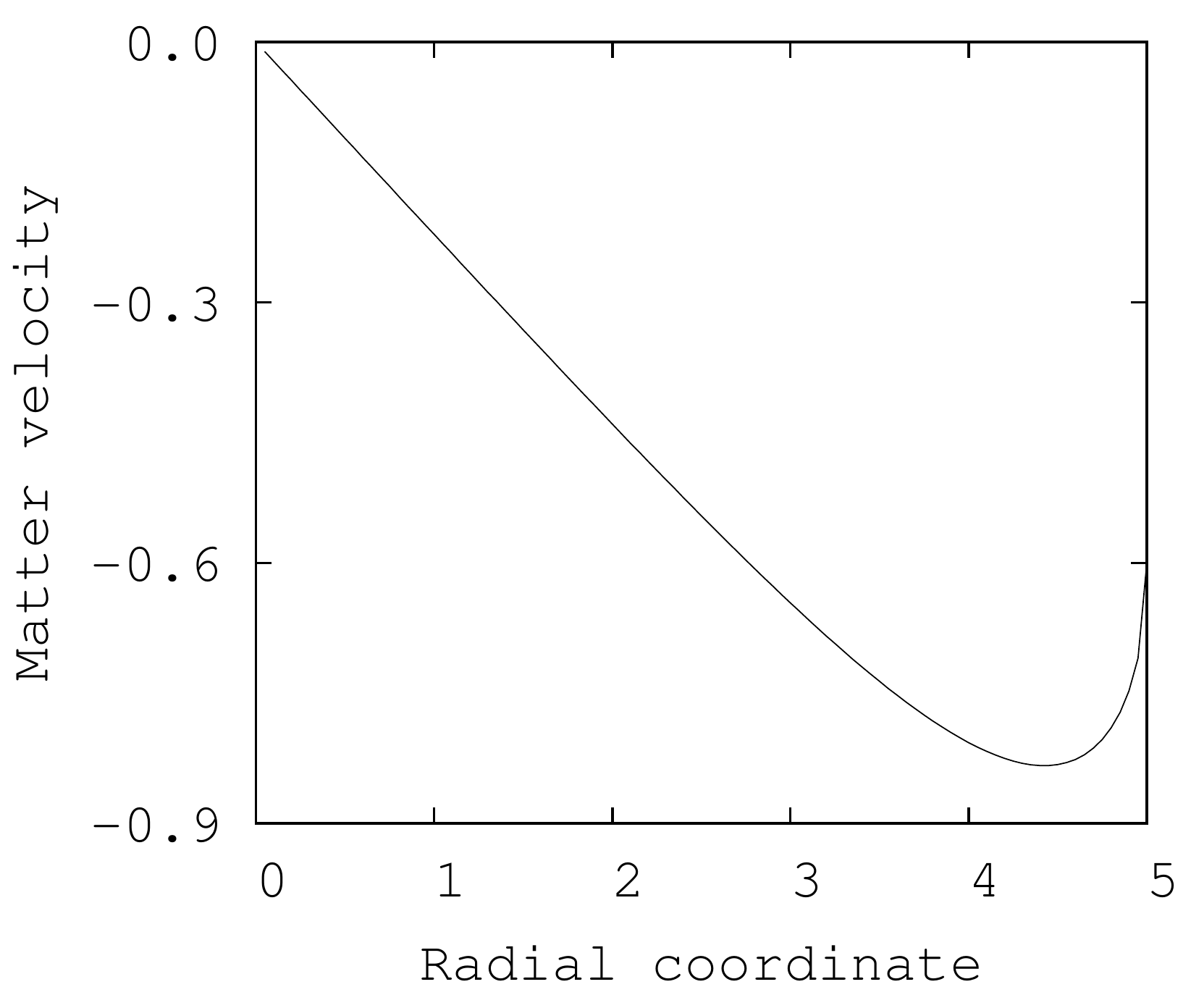}}
\caption{Initial profile of the matter velocity $dr/dt$ (multiplied by $10^3$) for the Schwarzschild--like model I. 
The initial conditions are $a(0) = 5.0$, $m_a(0)=1.0$, $\omega_a(0)=-10^{-3}$,
with $\ell=0.5$, which corresponds to $h=0.8966$.}
\end{center}
\label{fig:figure5}
\end{figure}

\begin{figure}[htbp!]
\begin{center}
\scalebox{0.4}{\includegraphics[angle=0]{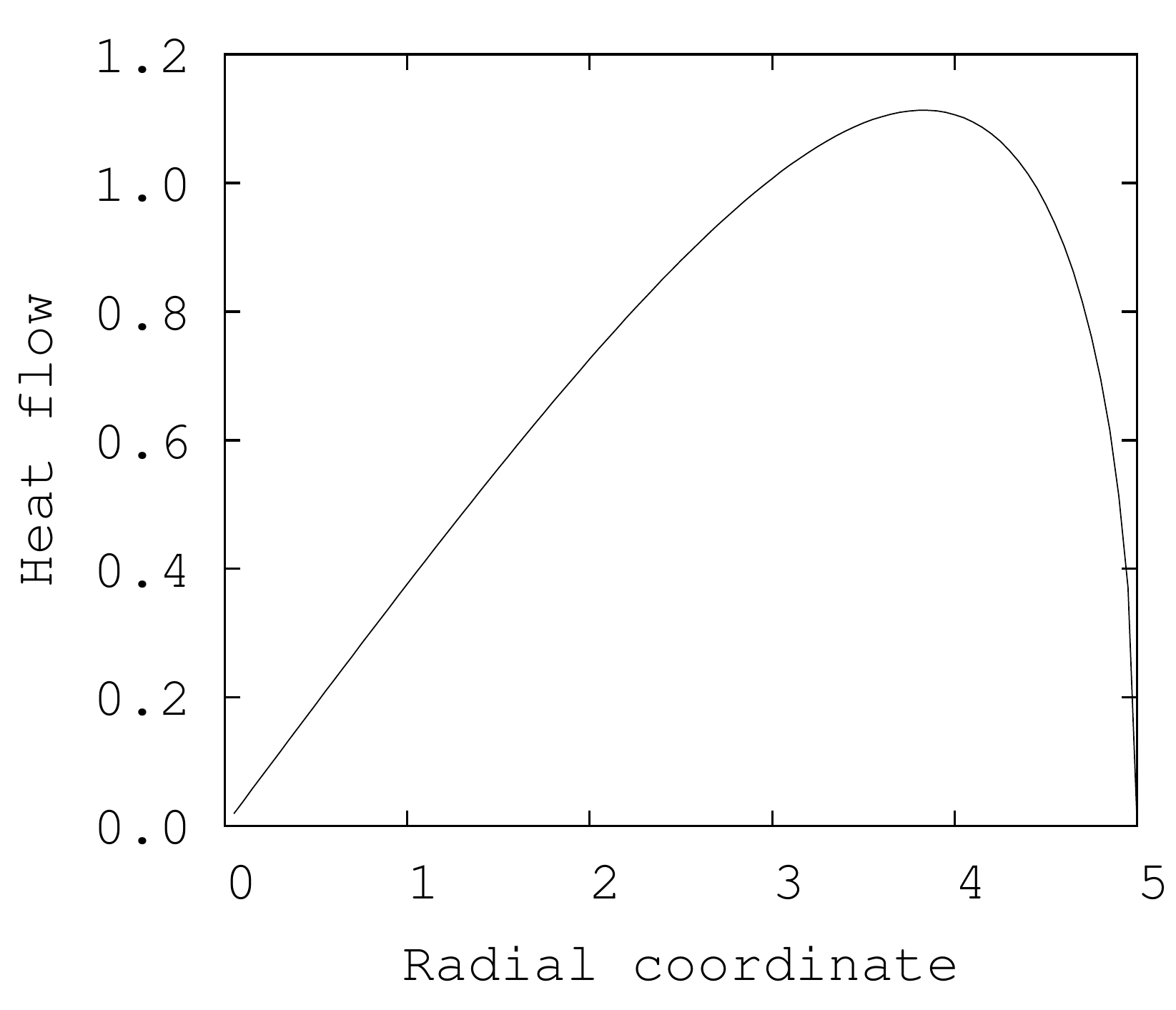}}
\caption{Initial profile of the heat flow $q$ (multiplied by $10^6$) for the Schwarzschild--like model I. 
The initial conditions are $a(0) = 5.0$, $m_a(0)=1.0$, $\omega_a(0)=-10^{-3}$,
with $\ell=0.5$, which corresponds to $h=0.8966$.}
\end{center}
\label{fig:figure6}
\end{figure}

\begin{figure}[htbp!]
\begin{center}
\scalebox{0.4}{\includegraphics[angle=0]{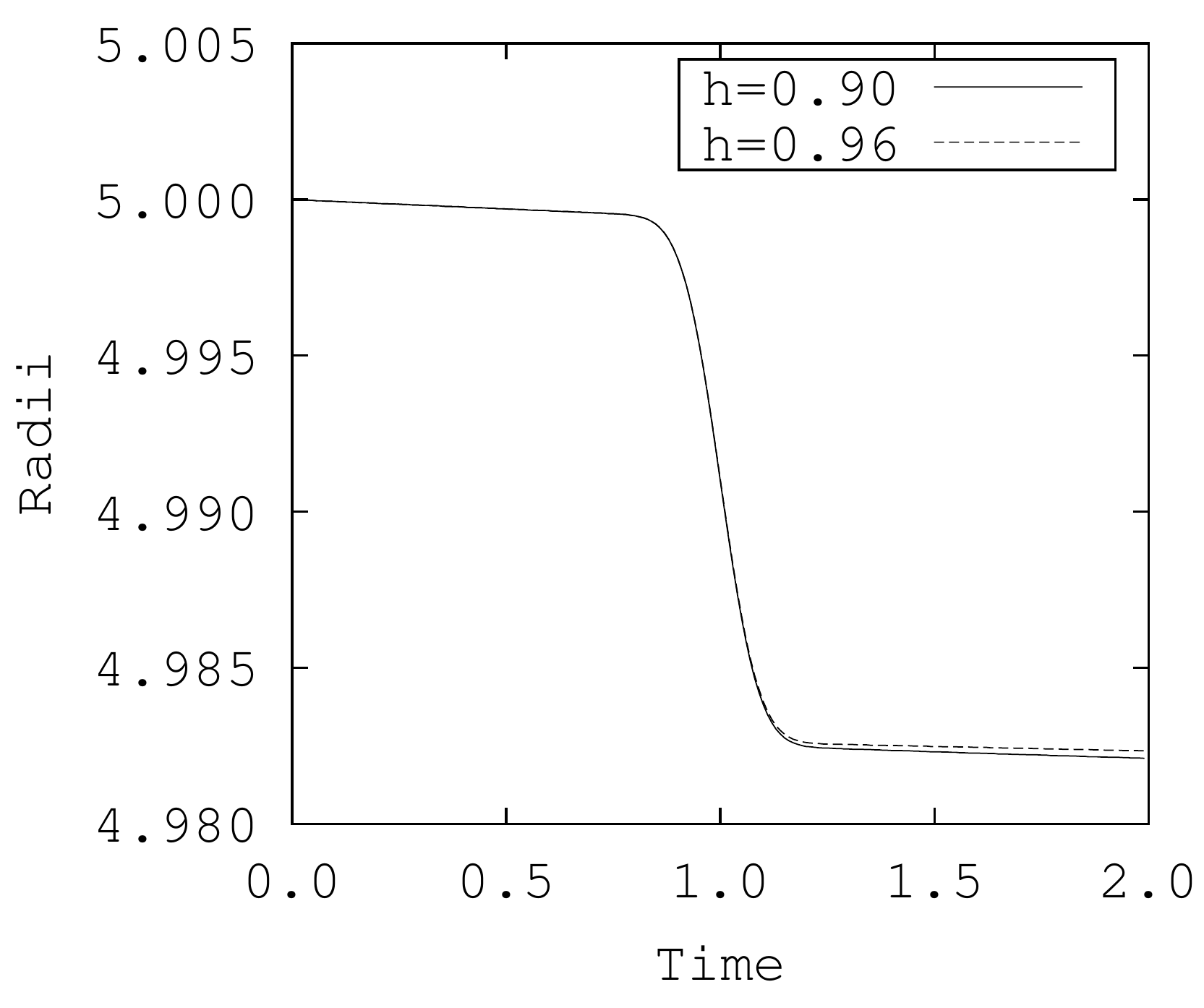}}
\caption{Evolution of radius $a$ for the Schwarzschild--like model I. 
The initial conditions are $a(0) = 5.0$, $m_a(0)=1.0$, $\omega_a(0)=-10^{-3}$,
with $h=0.90$ and $h=0.96$, corresponding to $\ell=0.491$ and $\ell=0.314$, respectively.}
\end{center}
\label{fig:figure7}
\end{figure}

\begin{figure}[htbp!]
\begin{center}
\scalebox{0.4}{\includegraphics[angle=0]{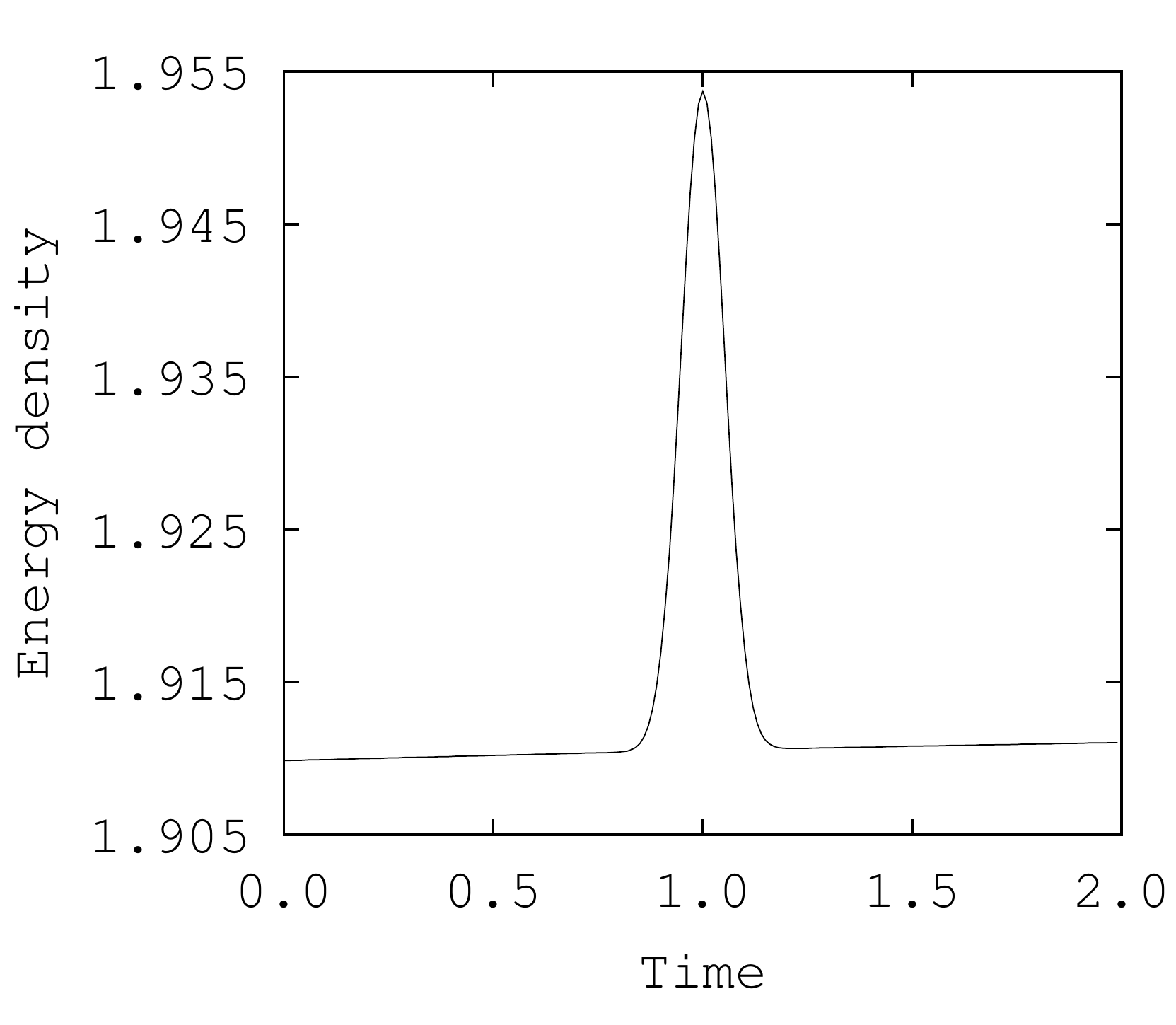}}
\caption{Evolution of the energy density $\rho$ (multiplied by $10^3$) at the surface for the Schwarzschild--like model I. The initial conditions are $a(0) = 5.0$, $m_a(0)=1.0$, $\omega_a(0)=-10^{-3}$,
with $\ell=0.5$, which corresponds to $h=0.8966$}
\end{center}
\label{fig:figure8}
\end{figure}

\begin{figure}[htbp!]
\begin{center}
\scalebox{0.4}{\includegraphics[angle=0]{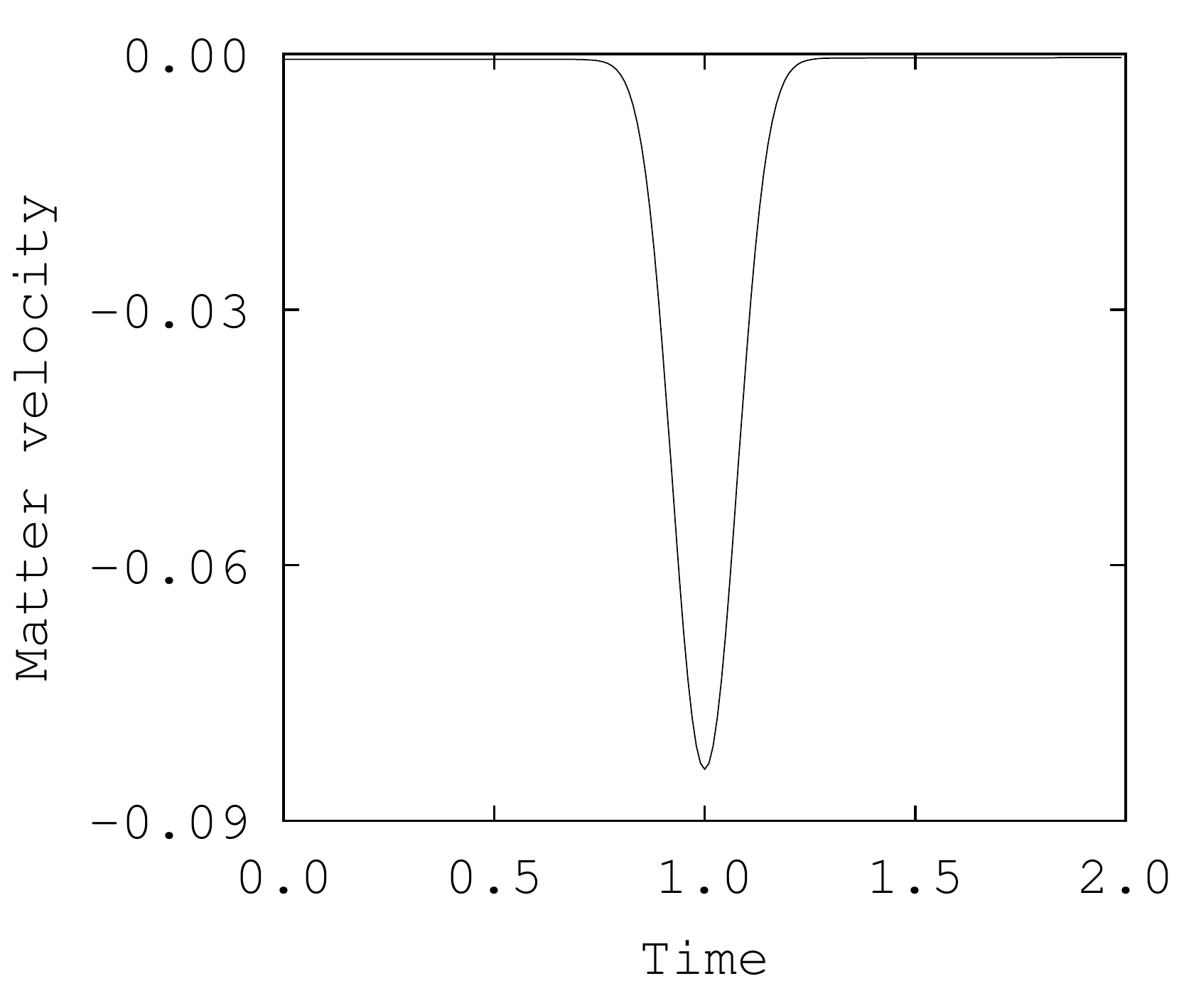}}
\caption{Evolution of the matter velocity $dr/dt$ at the surface for the Schwarzschild--like model I. The initial conditions are $a(0) = 5.0$, $m_a(0)=1.0$, $\omega_a(0)=-10^{-3}$,
with $\ell=0.5$, which corresponds to $h=0.8966$}
\end{center}
\label{fig:figure9}
\end{figure}

\begin{figure}[htbp!]
\begin{center}
\scalebox{0.4}{\includegraphics[angle=0]{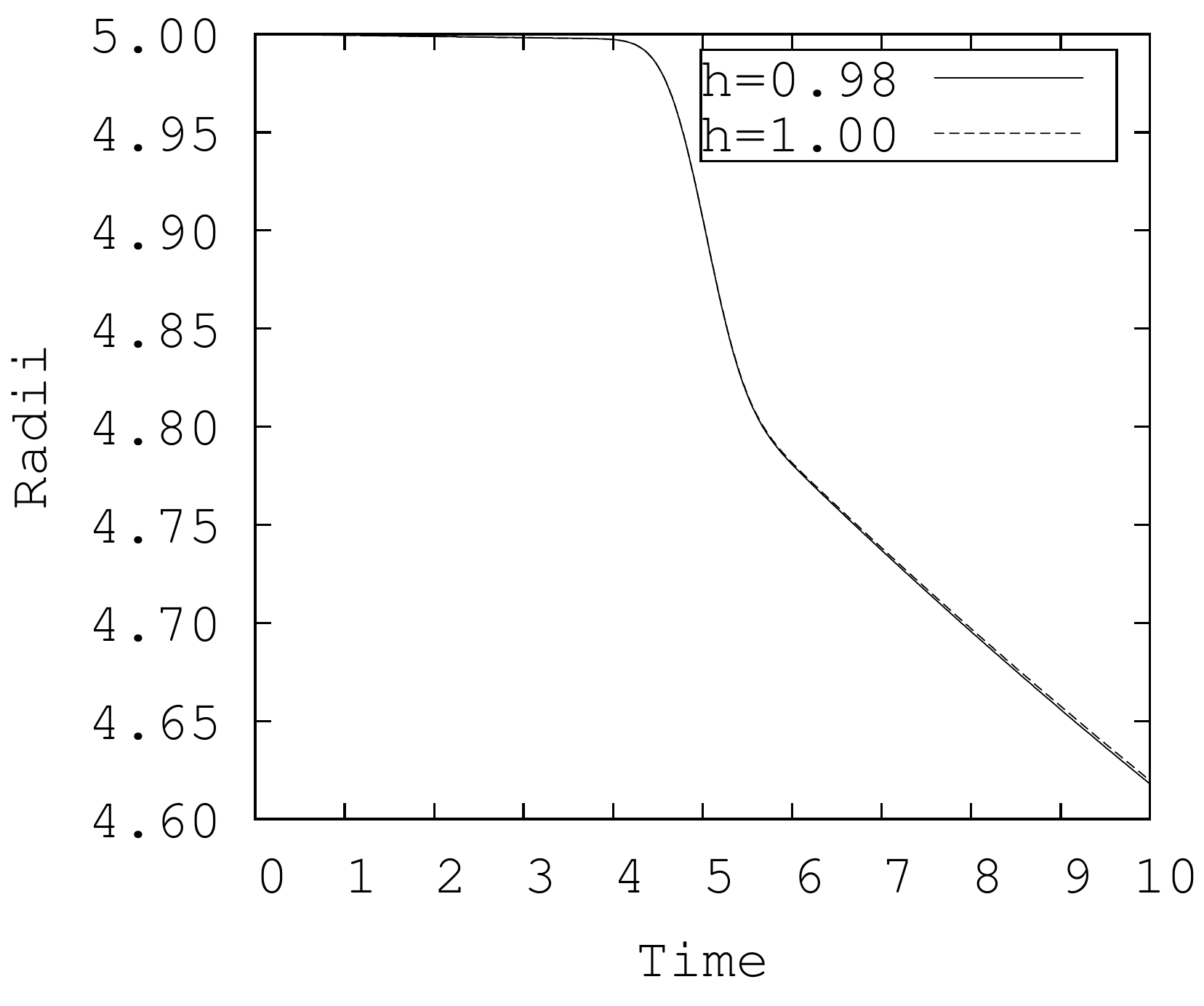}}
\caption{Evolution of radius $a$ for the Schwarzschild--like model II. 
The initial conditions are $a(0) = 5.0$, $m_a(0)=1.0$, $\omega_a(0)=-10^{-3}$,
with $h=0.98$ and $h=1$ corresponding to $\ell=0.222$ and $\ell=0$, respectively.}
\end{center}
\label{fig:figure10}
\end{figure}

\begin{figure}[htbp!]
\begin{center}
\scalebox{0.4}{\includegraphics[angle=0]{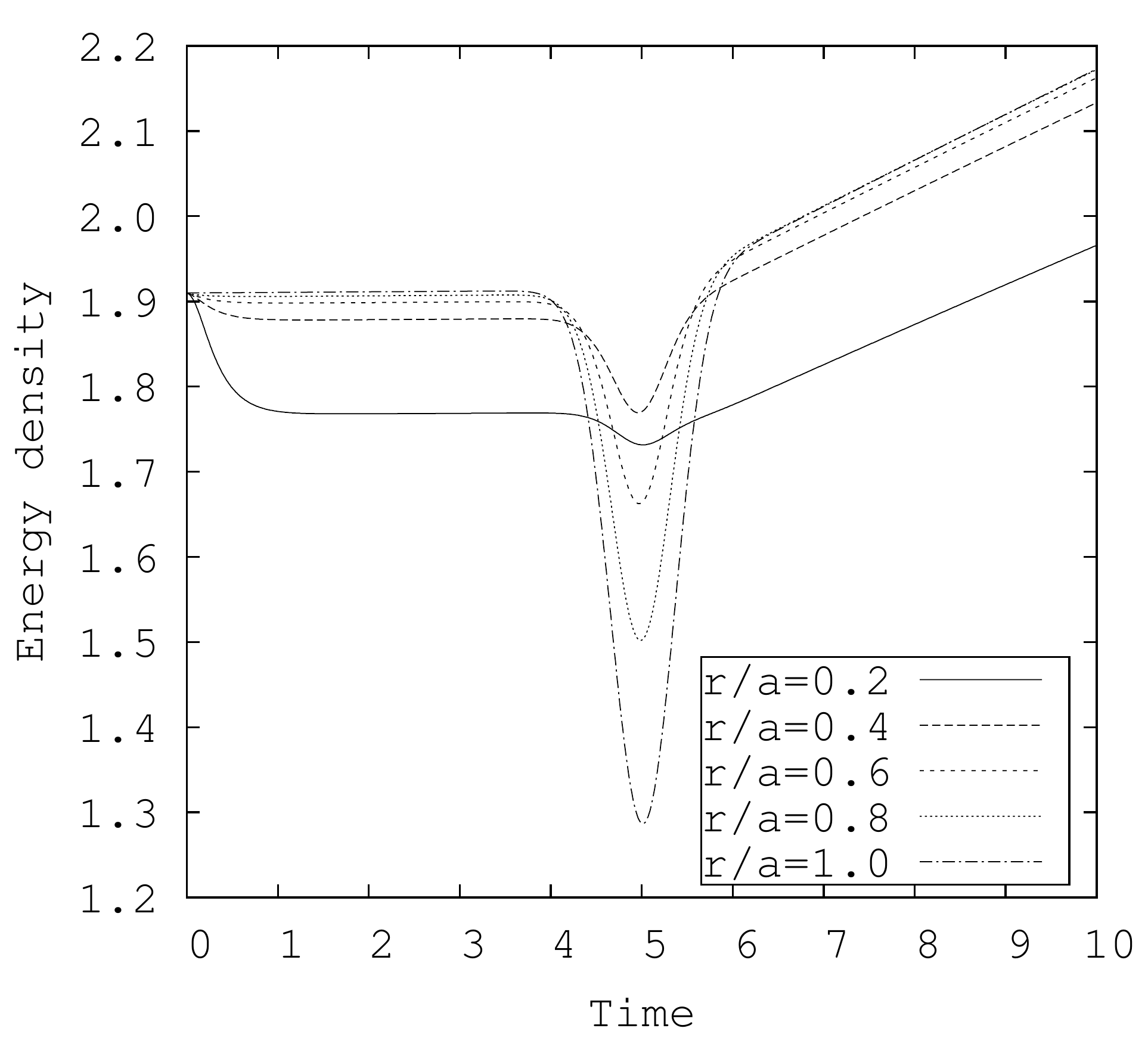}}
\caption{Evolution of the energy density $\rho$ (multiplied by $10^3$) for the Schwarzschild--like model II. 
The initial conditions are $a(0) = 5.0$, $m_a(0)=1.0$, $\omega_a(0)=-10^{-3}$,
with $\ell=0.2$, which corresponds to $h=0.9839$.}
\end{center}
\label{fig:figure11}
\end{figure}

\begin{figure}[htbp!]
\begin{center}
\scalebox{0.4}{\includegraphics[angle=0]{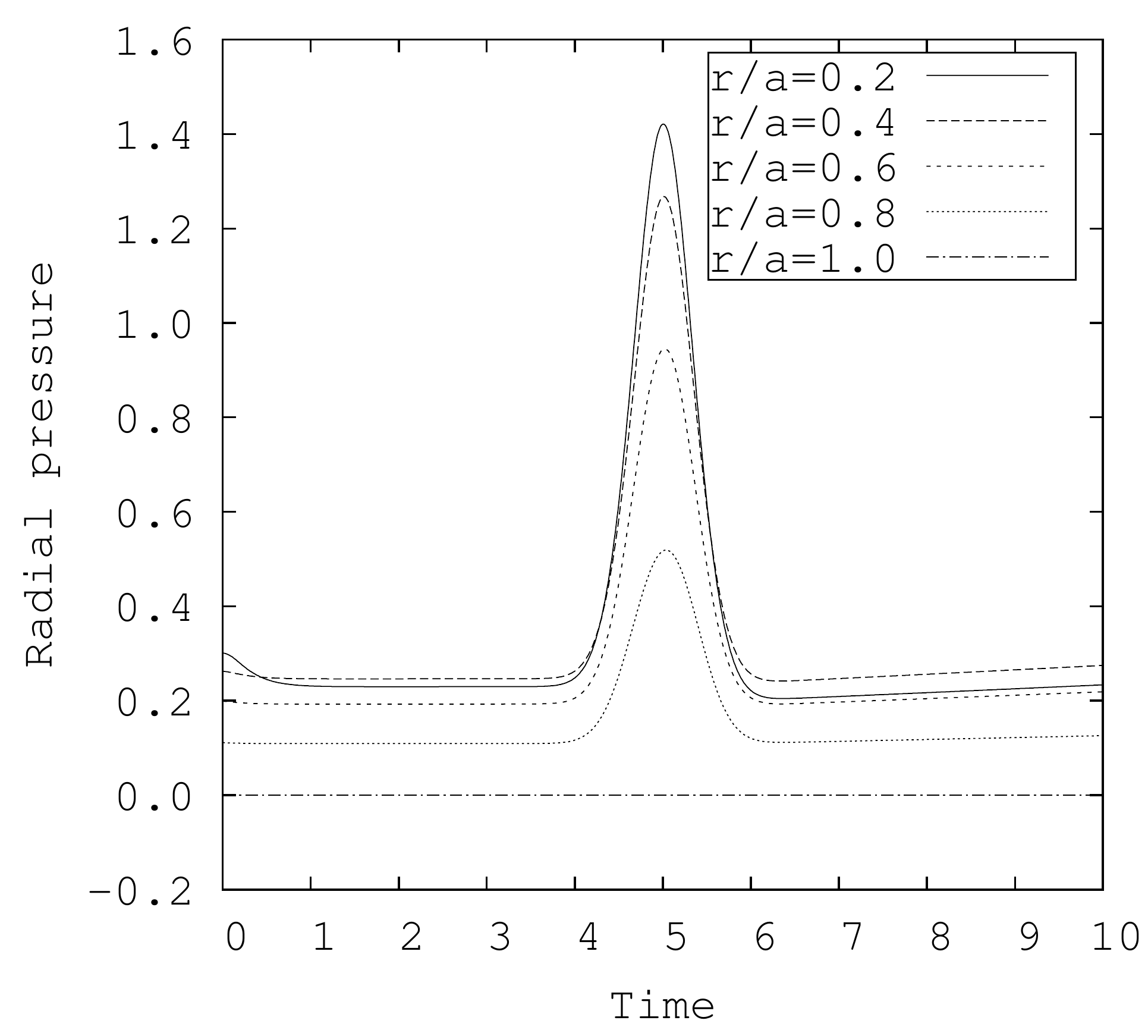}}
\caption{Evolution of the radial pressure $p$ (multiplied by $10^3$) for the Schwarzschild--like model II. 
The initial conditions are $a(0) = 5.0$, $m_a(0)=1.0$, $\omega_a(0)=-10^{-3}$,
with $\ell=0.2$, which corresponds to $h=0.9839$.}
\end{center}
\label{fig:figure12}
\end{figure}

\begin{figure}[htbp!]
\begin{center}
\scalebox{0.4}{\includegraphics[angle=0]{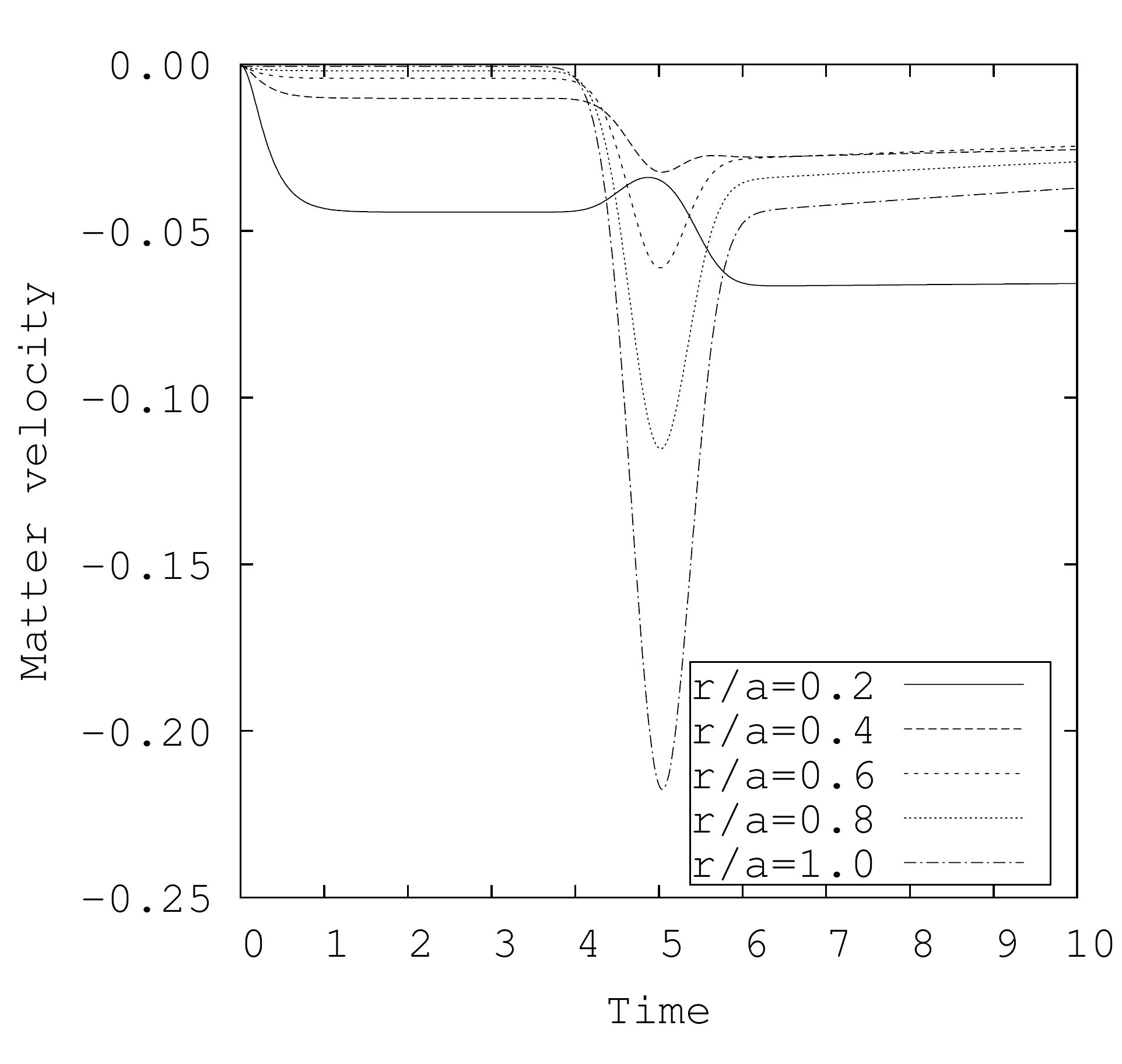}}
\caption{Evolution of the matter velocity $dr/dt$ for the Schwarzschild--like model II. 
The initial conditions are $a(0) = 5.0$, $m_a(0)=1.0$, $\omega_a(0)=-10^{-3}$,
with $\ell=0.2$, which corresponds to $h=0.9839$.}
\end{center}
\label{fig:figure13}
\end{figure}

\begin{figure}[htbp!]
\begin{center}
\scalebox{0.4}{\includegraphics[angle=0]{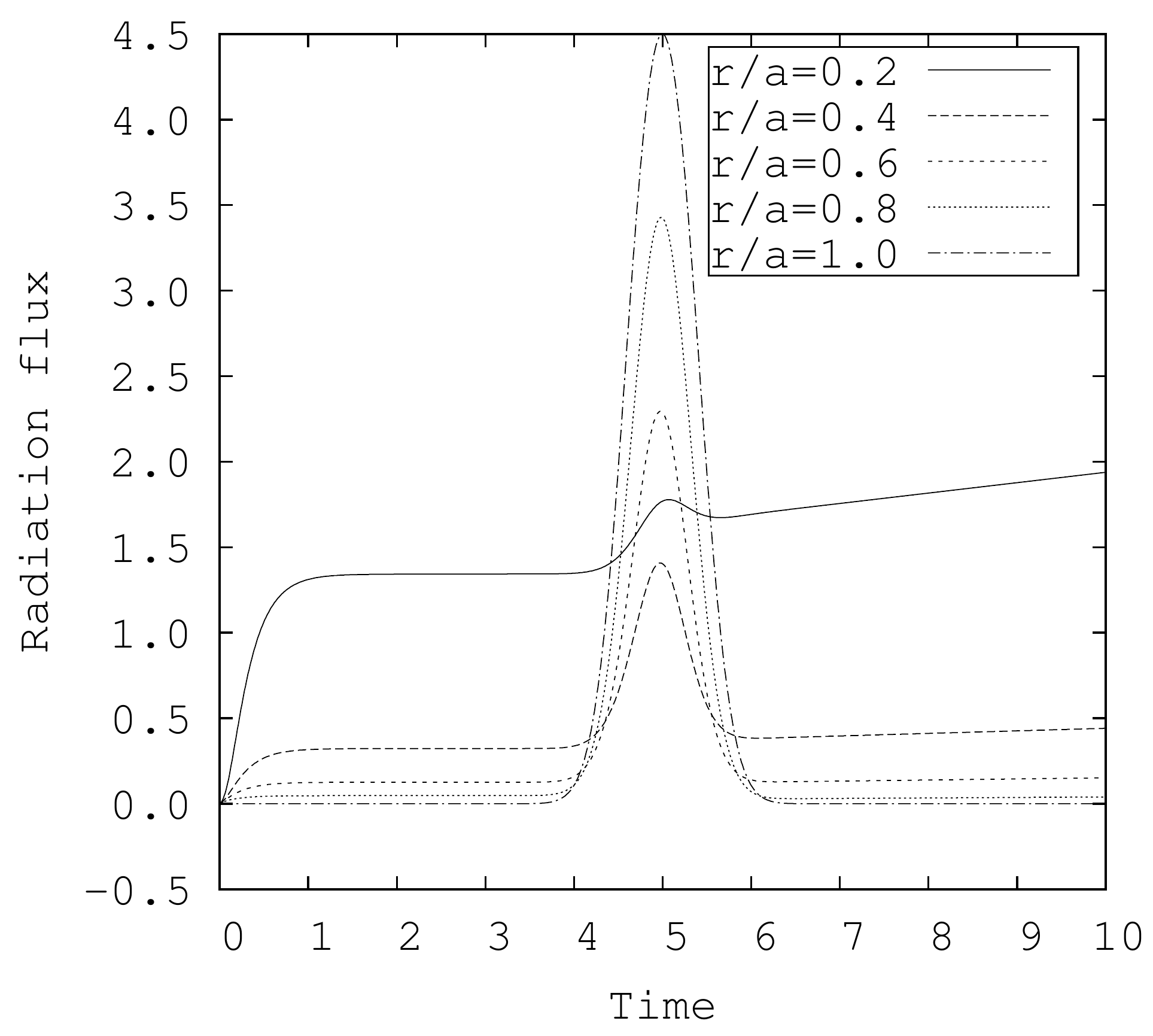}}
\caption{Evolution of the radiation flux $\epsilon$ (multiplied by $10^4$) for the Schwarzschild--like model II. 
The initial conditions are $a(0) = 5.0$, $m_a(0)=1.0$, $\omega_a(0)=-10^{-3}$,
with $\ell=0.2$, which corresponds to $h=0.9839$.}
\end{center}
\label{fig:figure14}
\end{figure}

\begin{figure}[htbp!]
\begin{center}
\scalebox{0.4}{\includegraphics[angle=0]{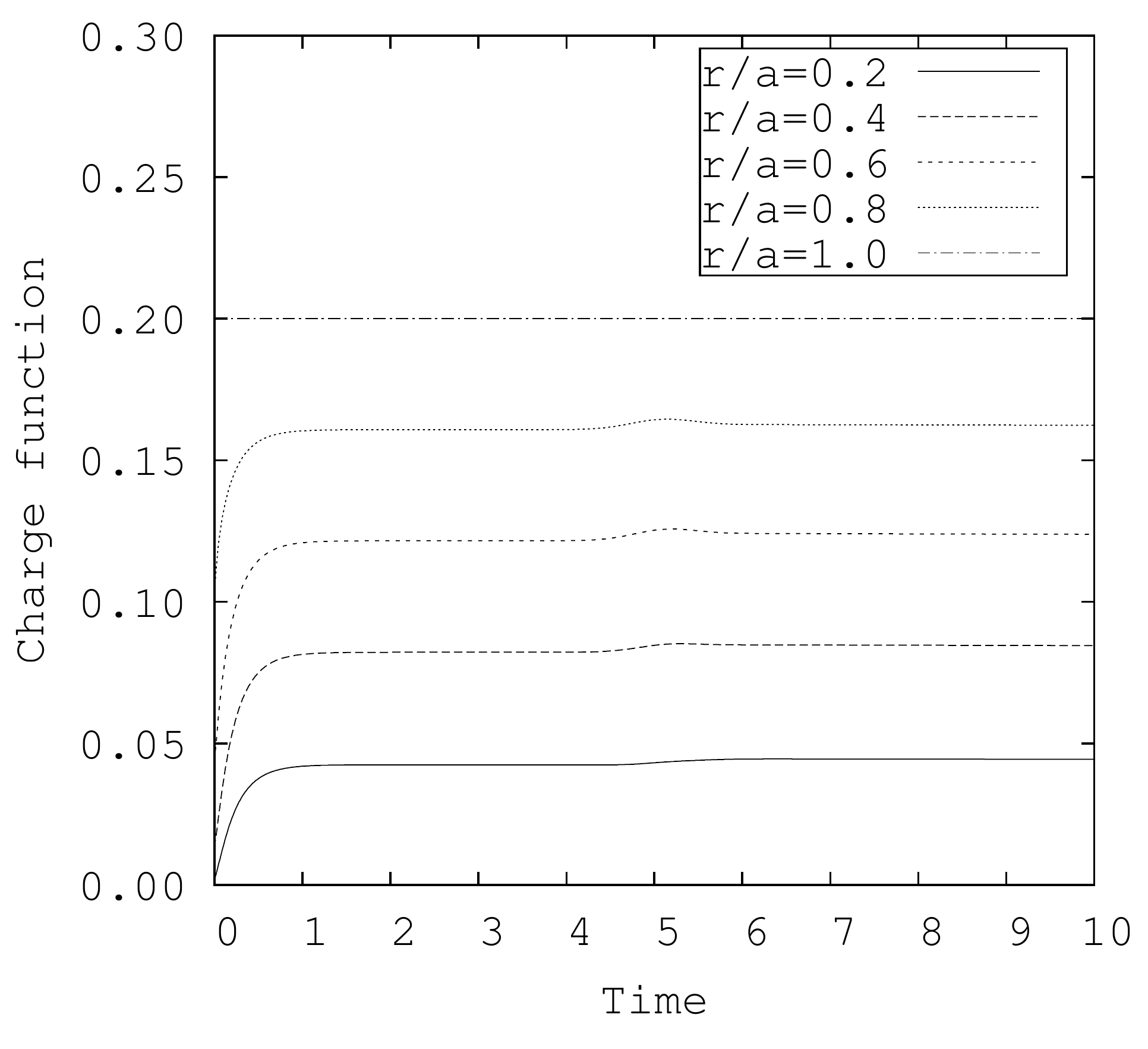}}
\caption{Evolution of the charge function $s$ for the Schwarzschild--like model II. 
The initial conditions are $a(0) = 5.0$, $m_a(0)=1.0$, $\omega_a(0)=-10^{-3}$,
with $\ell=0.2$, which corresponds to $h=0.9839$.}
\end{center}
\label{fig:figure15}
\end{figure}

\begin{figure}[htbp!]
\begin{center}
\scalebox{0.4}{\includegraphics[angle=0]{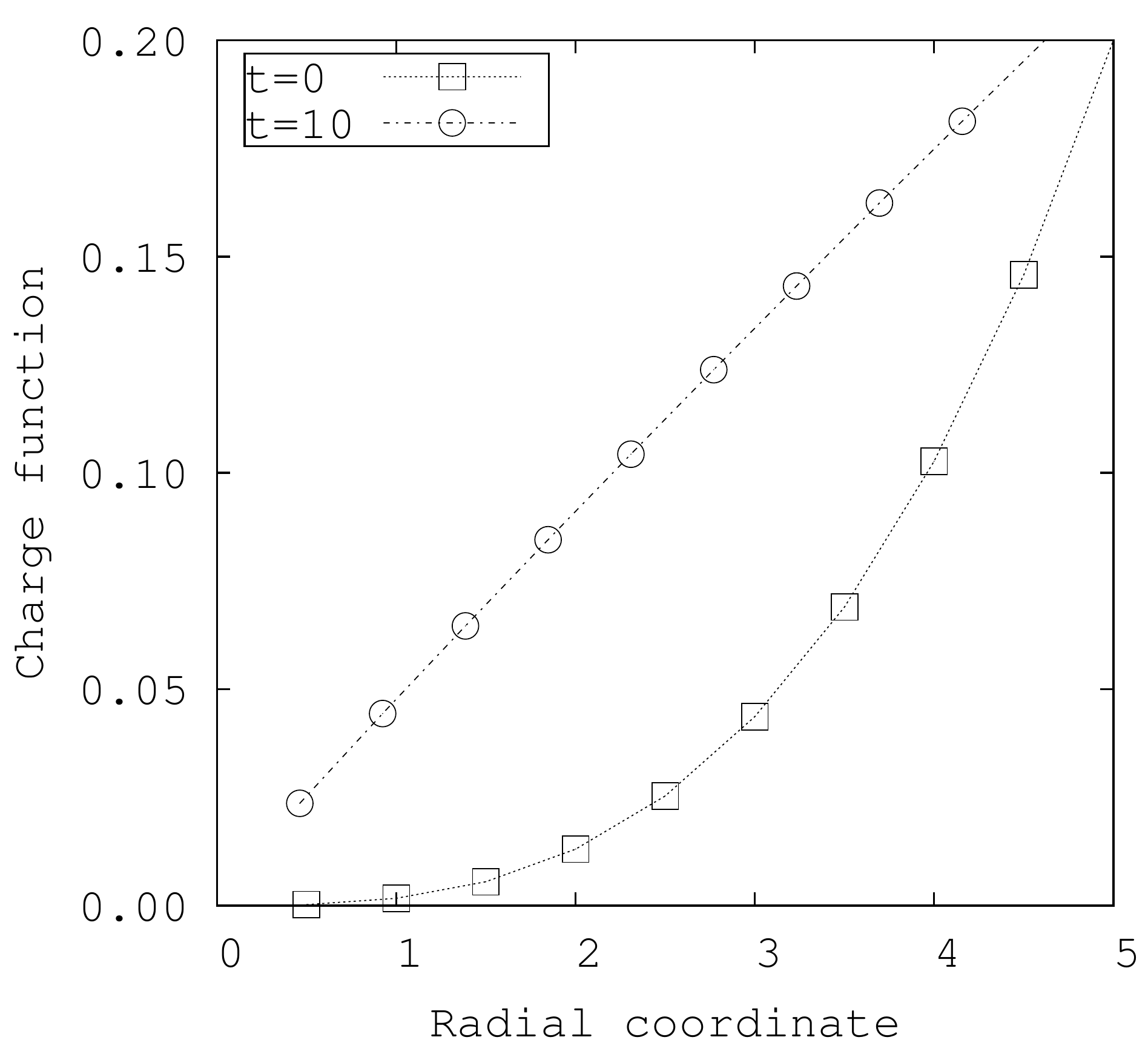}}
\caption{Initial and final charge function $s$ profile for the Schwarzschild--like model II. 
The initial conditions are $a(0) = 5.0$, $m_a(0)=1.0$, $\omega_a(0)=-10^{-3}$,
with $\ell=0.2$, which corresponds to $h=0.9839$.}
\end{center}
\label{fig:figure16}
\end{figure}

{\section{Numerical methods}}
{Once the surface equations are integrated using a standard method, as Runge--Kutta of 4th order (RK4),
we have to integrate the conservation equation (\ref{eq:ce}) to obtain all the physical
variables inside the source. 
Thus the conservation equation
\begin{equation}
s_{,t}=-\frac{dr}{dt} s_{,r}, \label{eq:ce2}
\end{equation}
is a wave--like equation that can be integrated numerically
using the Lax method (with the appropriate Courant--Friedrichs--Levy (CFL)
condition). The evolution of the conservation equation is restricted
by the surface evolution and is implemented as follows
\begin{equation}
s^{n+1}_j=\frac{1}{2}\left(s^n_{j+1}+s^n_{j-1}\right)-
\frac{\delta t}{2\delta r} \left(\frac{dr}{dt}\right)^n_j
\left(s^n_{j+1}-s^n_{j-1}\right). \label{eq:lax}
\end{equation}
The superscript $n$ indicates the hypersurface $t=n\delta t$
and the subscript $j$ the spatial position for a comoving
observer at $r=j\delta r$. In order to integrate
the conservation equation we need to specify a boundary--initial condition.}

The PQSA is a seminumerical method where the radial dependence
is determined from a static interior solution
and kept the same during the evolution. The problem
is typically  reduced  to a system of ordinary
differential equations (ODEs) at the surface
of the distribution of matter. This system is integrated
in time using the RK4
method. Therefore we can calculate exactly any physical variable
at the interior. In our specific case,
for the Einstein--Maxwell system, the approach requires
additionally the integration of the conservation of the electric charge
(Eq. (\ref{eq:ce2})) using the Lax  method (Eq. (\ref{eq:lax})). 
The conservation equation is an evolution equation
constrained by the system of ODEs at the surface.
Thus, the numerical convergence of the
whole algorithm must be of 2nd order accuracy.

The implemented algorithm at the surface (basically that of RK4) for
the specific models
was verified with satisfaction from a physical point of view and within
a reasonable numerical error (see subsection IV.A). If the numerical solution 
for the electric charge function is stable and globally convergent to 2nd order, as
shown in subsection IV.B, the problem surely is well--posed \cite{referee}. 

\section{Testing and Modeling}
To illustrate the method let us consider a Schwarzschild--like model. Following the protocol for the PQSA \cite{hbds02}, \cite{prrb10} the interior solution has the effective density
\begin{equation}
\tilde\rho=f(t),
\end{equation}
where $f$ is an arbitrary function of time. 
The expression for $\tilde p$ is
\begin{equation}
\frac{\tilde p + \frac{1}{3}\tilde\rho}{\tilde p + \tilde\rho}=
\left(1-\frac{8\pi}{3}\tilde\rho r^2 \right)^{h/2}k(t), \label{eq:prep}
\end{equation}
where $k$ is a function of $t$ to be defined from the boundary condition
(\ref{eq:boundary}), which now reads, in terms of the effective variables, as
 \begin{equation}
\tilde p_a=\tilde\rho_a\omega_a^2 + (q_a+\epsilon_a) (1 + \omega_a)^2-(1+\omega_a^2)\frac{\ell^2}{8\pi a^4}.
 \label{eq:boun}
\end{equation}
Thus, (\ref{eq:prep}) and (\ref{eq:boun}) give
\begin{equation}
\tilde\rho=\frac{3\mu_a}{4\pi a^3},
\end{equation}
\begin{equation}
\tilde p=\frac{\tilde\rho}{3}\Biggl\{\frac{\chi_S (1-2\mu_a/a)^{h/2} -3\psi_S\xi}
{\psi_S\xi -\chi_S (1-2\mu_a/a)^{h/2}}\Biggr\}, \label{eq:effepre}
\end{equation}
with
$$
\xi=\left[1-\frac{2\mu_a}{a}(r/a)^2\right]^{h/2}
$$
where $h=1-2C$
\begin{equation}\chi_S=6(\omega_a^2+1)\frac{\mu_a}{a}+2(Q+E)(1+\omega_a)^2
-(1+\omega_a^2)\frac{\ell^2}{a^2},
\end{equation}
and
\begin{equation}\psi_S=2(3\omega_a^2+1)\frac{\mu_a}{a}+2(Q+E)(1+\omega_a)^2
-(1+\omega_a^2)\frac{\ell^2}{a^2}.
\end{equation}
Using (\ref{eq:e1}) and (\ref{eq:e2}) it is easy to obtain expressions
for $\mu$ and $\nu$:
\begin{equation}
\mu=\mu_a(r/a)^3, \label{eq:mass_sch}
\end{equation}
\begin{equation}
e^{\nu}=\Biggl\{\frac{a[\chi_S (1-2\mu_a/a)^{h/2}-\psi_S\xi]}{4\mu_a}
\Biggr\}^{2/h},\\ \label{eq:nu_sch}
\end{equation}
which correspond to the Hamiltonian constraint and
to the polar slicing condition  in the ADM 3+1 formulation, respectively.
It is necessary to specify one function of $t$ and the initial data. 

We choose 
$L(t)$ to be a Gaussian  
\begin{equation}
L=L_0 e^{-(t-t_0)^2/\Sigma^2},
\end{equation}
with $L_0=M_r/\sqrt{\Sigma \pi}$, which corresponds to a pulse
radiating away a fraction of the initial mass $M_r$.
It is easy to construct the initial profile for the charge function. From (\ref{eq:EDE}) evaluated at the surface we obtain $h(\ell)$ (see Figure 1). Figure 2 displays the charge function $s$ for the following set of initial conditions
\begin{equation}
a(0)=5,\;\;\;m_a(0)=1,\;\; \omega_a(0)=-10^{-3},
\end{equation}
with $\ell=0.5$, which corresponds to $h=0.8966$.

\subsection{Model I}

In the diffusion limit ($\epsilon=0$) we choose $t_0=1.0$ and $\Sigma=0.01$, $M_r=10^{-2}$.
We tested that the algorithm is correct at the surface (that is, locally) by verifying
that the pressure is equal
to the given gaussian pulse with an accuracy
of about $10^{-19}$. This allow us at least to be confident
about the implemented algorithm at the surface.
However, as a double--check, from an strictly numerical
point of view,  we show in Table I a proper convergence test
to 4th order of the RK algorithm at the surface.
For this test we use the gravitational potential at the
surface as the required norm, $\mathcal{N}=\mu_a/a$. Thus,
it can be shown that the rate of convergence is
\begin{equation}
n=\log_2\frac{\mathcal{N}_c-\mathcal{N}_m}{{\mathcal N}_m-{\mathcal N}_f} \label{eq:formula},
\end{equation}
where $\mathcal{N}_c$, $\mathcal{N}_m$ and $\mathcal{N}_f$ are values 
of $\mathcal{N}$ for a coarse, medium, and fine time step $\Delta t$,
respectively (scaling as 4:2:1, \cite{bghlw03},\cite{bdglrw05}). This corresponds
to a local convergence test for the RK4, giving a convergence rate $n \approx 4$, as expected.

$\;$
\begin{table}[!ht]
\begin{tabular}{|l|c|c|c|}
\hline
$t$&$\mathcal{N}_c-\mathcal{N}_m \,(10^{-10})$&$\mathcal{N}_m-\mathcal{N}_f \,(10^{-12})$ &$n$\\
\hline
$1.0$  & -0.022 & -0.142 &$3.964$\\
\hline
$1.5$  & -0.573& -3.586 &$3.998$\\
\hline
$2.0$  & -1.148& -7.180 &$3.999$\\
\hline
\end{tabular}
\caption{Proper convergence of the surface gravitational potential;
the expected value for $n$ is $4$.}
\label{tab:expansion}
\end{table}

For the initial setting corresponding to Figure 2 the interior does not evolve
because the velocity becomes a complex value. However, we can display
for analysis the initial set of physical variables in Figures 3--6.
Under these conditions the surface evolves without anomalies 
as shown in Figures 7--9. We have looked for an electrically
charged initial configuration for which all regions evolve; we found one with a total electric  charge $\approx 10^{-6}$. 

\subsection{Model II}
Let us consider now a Schwarzschild--like model in the streaming--out limit ($q =0$).
We choose $t_0=5.0$ and $\Sigma=0.25$, $M_r=10^{-1}$. We tested again that the algorithm is correct at the surface by verifying
that the pressure is equal to zero with an accuracy of about $10^{-19}$.
For this model we show the global rate of proper convergence in Table II. For that
purpose we construct the following norm with the electric charge function $s$
\begin{equation}
\mathcal{N}=\int^a_0 s^2 dr. \label{eq:norm}
\end{equation}
$\;$
\begin{table}[!ht]
\begin{tabular}{|l|c|c|c|c|}
\hline
$t \,(10^{-7})$&$\mathcal{N}_c \,(10^{-3})$&$\mathcal{N}_m \,(10^{-3})$ &$\mathcal{N}_f \,(10^{-3})$ &$n$\\
\hline
\;\;\;\;0  & 3.5423 & 3.5322 & 3.5297 & 2.0152\\
\hline
\;\;\;\;1  & 5.1726 & 5.1564 & 5.1524 & 2.0398\\
\hline
\end{tabular}
\caption{Proper convergence of the norm (\ref{eq:norm});
the expected value for $n$ is $2$.}
\label{tab:expansion}
\end{table}
For instance, we choose grids of 10, 20 and 40 nodes, and a RK4 time step of $10^{-8}$ in a proportion of 4:2:1, respectively. To reach the same monitoring time the CFL time step have to
be $\delta t = K \delta r$, where $K$ is a constant of order of one,
in proportion 4:2:1 as well. Consequently, the number of RK4 time steps required
to apply the formula (\ref{eq:formula}) is in proportion 1:4:16. The global convergence
test (RK4 + Lax) gives a rate of $n\approx 2$, as expected.

The initial setting shown in Figure 2 is valid for this model too, but with $\ell=0.2$ (corresponding to $h=0.8966$). A larger total electric charge does not allow an interior evolution because unphysical values develop.
Figures 10--16 show these results. 
The collapse is unavoidable after emitting an energy
equivalent to 10 \% of the initial mass,
which decreases the energy density (the contrary occurs
in the diffusion limit \cite{rprb10}). The electric charge is redistributed in the whole body. 
Note that the electric charge on each comoving shell starts moving
until it reaches a stationary state while the whole distribution collapses. 
Observe how the gradient of charge decreases and becomes linear with the advance of time.
\section{Concluding remarks}
\label{sec:conc}
We consider the evolution of a self--gravitating spherical distribution of charged matter containing a dissipative fluid. The use of the PQSA with noncomoving coordinates
allows us to study electrically charged fluid spheres in the diffusion and the streaming out limits as they just depart from equilibrium. From this point of view,
the PQSA can also be seen as a nonlinear perturbative method 
to test the stability of solutions in equilibrium.
{We have shown that our seminumerical implementation is globally second--order convergent.}

Our results indicate that the dissipative transport
mechanisms and the equation of state chosen to treat
electric charge as anisotropy are crucial for the outcome of
gravitational collapse. We want to stress: i) the straightforward manner
in which we connected anisotropy with electric charge
using an EoS; ii) how the EoS is used in practice as an
initial--boundary condition; iii) that departing from the
same static solutions we find very different evolutions.
Increasing the amount of total charge $\ell$ results in
lower values of the anisotropy parameter $h$, approaching zero. 
The zero limit is not reached because the system breaks down 
for some limit value of the total electric charge. When the
transport mechanism is diffusive, it was not possible
to establish why the system can be set initially but not evolved,
except at the surface. 
From the results, the field equations seem to be imposing restrictions within the context of 
diffusion and electric charge (or anisotropy) and permit only bubbles of charged matter.
Otherwise the electric
charge (or anisotropy) has to be very small. In the streaming out
limit the situation is quite different, as is expected. 
The interior is evolved for a total charge that is 200,000 times that
of the maximum charge permitted in the diffusion limit.  
Coupling of matter with radiation is not strong enough to prevent the
collapse and the system efficiently radiates a large quantity of mass. 
The system clearly departs equilibrium and collapses.
Electric charge contributes
to the collapse in the same way that anisotropy with tangential
pressure greater than radial pressure favors the collapse \cite{prrb10},
irrespective of the transport mechanism. In any case electric charge
has to be huge to change the fate of the gravitational collapse.
The electric charge is redistributed in a such way that its gradient
decreases toward the surface and becomes unexpectedly linear
and stationary, with the advance of time. 
There is a critical total electric
charge (or anisotropy parameter) for which the system evolves
constrained by the Einstein--Maxwell system of field equations.

Beyond the models we want to stress some features about 
our framework. First, the luminosity profiles are given as Gaussian
but they can be provided from observational data. 
To keep physical variables on appropriate
values in the diffusion approximation, the pulse has to
be narrow in comparison, while the streaming out limit
allows for a wider pulse.
Second, the EoS used in this work is not essential. Other EoS as initial--boundary
data should fit well, understanding that it represents anisotropic
matter. Third, from the observational point of view, temperature
profiles are desirable as input data, but they are not available
in the PQSA method when electric charge is taked in account.   

We considered the dissipation by viscosity and
heat flow separately \cite{prrb10}, \cite{rprb10}, in order  to isolate similar effects with different
mechanisms. In this work we considered heat flow/streaming out
and anisotropy induced by electric charge,  
pointing to the most realistic numerical modeling in
this area  \cite{dhlms07}. 
The results constitute a definite first cut to more general situations
using the PQSA, including dissipation, anisotropy, electric charge, heat flow,
viscosity, radiation flux, superficial tension, temperature profiles
and study their influence on the gravitational collapse.
{Numerical issues apart, the inclusion of superficial tension \cite{hjei89} together with a highly compressed Fermi gas \cite{hbds02}, and more realistic thermal processes \cite{hdb06}, is of current interest in astrophysics \cite{dhlms07}. Cooling times of smooth or crusty surfaces may be the way to differentiate strange stars from neutron stars \cite{jrs06}. The PQSA can be used to model these situations.}
This investigation is an essential part of a long--term project which 
tries to incorporate the M\"uller--Israel--Stewart theory for dissipation and deviations from spherical symmetry, specially when considering electrically charged distributions. Besides being interesting in their own right, we believe that spherically symmetric fluid models are useful as a test bed for more general solvers 
in numerical relativity \cite{bcb09,bcb10}. A general three--dimensional code must also be able
to reproduce situations closer to equilibrium. 

\begin{acknowledgments}
C. P. acknowledges the computing resources provided by the Victorian Partnership for Advanced Computation (VPAC).
\end{acknowledgments}

\end{document}